\documentclass[aps,9pt,twocolumn,showpacs,showkeys]{revtex4}
\usepackage[dvips]{graphicx}

\begin{document}
\title{On the crystal field in the modern solid-state theory$^\spadesuit$}

\author{R. J. Radwanski}
\homepage{http://www.css-physics.edu.pl}
\email{sfradwan@cyf-kr.edu.pl}
\affiliation{Center of Solid State Physics, S$^{nt}$Filip 5, 31-150 Krakow, Poland,\\
Institute of Physics, Pedagogical University, 30-084 Krakow, Poland}
\author{Z. Ropka}
\affiliation{Center of Solid State Physics, S$^{nt}$Filip 5,
31-150 Krakow, Poland}

\begin{abstract}
We point out the high physical correctness of the use and the
concept of the crystal-field approach, even if is used to metallic
magnetic materials of transition-metal 3d/4f/5f compounds. We
discuss the place of the crystal-field theory in modern
solid-state physics and we point out the necessity to consider
the crystal-field approach with the spin-orbit coupling and
strong electron correlations, as a contrast to the
single-electron version of the crystal field customarily used for
3d electrons. We have extended the strongly-correlated
crystal-field theory to a Quantum Atomistic Solid-State Theory
(QUASST) to account for the translational symmetry and inter-site
spin-dependent interactions indispensable for formation of
magnetically-ordered state. We have correlated macroscopic
magnetic and electronic properties with the atomic-scale
electronic structure for ErNi$_5$, UPd$_2$Al$_3$, FeBr$_2$,
LaCoO$_3$ and LaMnO$_3$. In QUASST we have made unification of 3d
and rare-earth compounds in description of the low-energy
electronic structures and magnetism of open 3d-/4f-/5f-shell
electrons. QUASST offers consistent description of
zero-temperature properties and thermodynamic properties of
4f-/5f-/3d-atom containing compounds. Our studies indicate that
it is the highest time to unquench the orbital magnetism in 3d
oxides.

\end{abstract}
\maketitle\vspace {-1.0cm}

\section{Introduction}
\vspace {-0.5cm}The use and the concept of the crystal-field
approach, if used to metallic magnetic materials, has been
recognized as erroneous by the Highest Scientific Council of the
Polish Government (CK ds SiTN, later in short CK) in its decision
BCK-V-O-819/03 on 31.05.2004. The full decision, in Polish and in
part in English, is available on www.css-physics.edu.pl.. This
reproach we denote as No 1. Such a formulation of the reproach
should not be read that this Council agrees that the use and the
concept of the crystal-field approach, if used to nonmetallic
magnetic materials, is correct. The above decision is somehow in
a common line of depreciation of the crystal field (CEF) theory in
the modern solid-state theories. Let mention, that a recently
edited (2003) book of Mohn " Magnetism in the solid state" [1],
being an overview of presently-in-fashion magnetic theories,
mentions only one position on the localized magnetism: a book of
Van Vleck from 1932 [2] (apart of this shortage the book is very
nice and we highly recommend it). Diagrams of Tanabe-Sugano [3],
known already 50 years, are not exploited in the modern
solid-state physics theories for description of 3d-ion compounds
and the orbital magnetic moment only recently starts to draw the
proper attention. Theoretical approaches yielding continuous wide
1-10 eV energy bands for 3d/4f/5f states overwhelm the present
solid-state theory. On other side the CEF approach, yielding the
discrete electronic structure for 3d/4f/5f electrons with details
at least 1000 times smaller, below 1 meV, is often recalled by
experimentalists in order to analyze obtained experimental
results. Thus, one can say that there is at present a large gap
between theory and experiment in description of 3d/4f/5f states.

In this paper we would like to discuss the place of the
crystal-field theory in the modern solid-state physics, to
clarify our understanding of the crystal field approach and to
inform about the administrative interference to Physics  in
judging the physical correctness, rather incorrectness, of the
crystal-field theory to metallic magnetic materials. We claim
that the simplest and most natural theoretical approach, as the
CEF theory is, has not been exploited enough for showing its
physical adequacy and its total theoretical rejection is
premature.

We by years openly formulate the need of taking into account the
crystal field in description of 3d/4f/5f compounds, even these
exhibiting the heavy-fermion phenomena (mainly 4f or 5f
compounds) and insulating 3d oxides. In order to avoid undeserved
critics we do not claim that CEF explains everything but we claim
that CEF effects should be clarified at first (properly!!!) in
any analysis of physical properties of any transition-metal
compound. Also we do not claim to invent the crystal-field theory
- we point out its importance in the specific social conditions
of the end of the XX and the beginning of the XXI century, when
the crystal-field theory is somehow prohibited and rejected from
magnetic theories.

The biggest problem in this discussion is related to a fact that
the crystal-field theory has within the magnetic community in
last 30 years a special place - being continuously rejected from
the scientific life permanently appears as an unavoided approach
for explanation of properties of real compounds. The crystal-field
theory is in the modern solid-state theory like an unwilling
child, 75 years old already.
\vspace {-0.3cm}
\section{Further administrative details}
\vspace {-0.3cm}The above mentioned objection was the only one
scientific reproach in the administrative decision CK-04 towards
the disqualification of the scientific activity of R. J.
Radwanski. These scientific achievements account for a date of 7
May 2001 126 publications in international journals from the SCI
philadelphia list and two promoted doctors. The full list of
these publications is available on www.css-physics.edu.pl.. There
are also further 19 publications to 31.12.2004. See also 51
internet papers in ArXiv/cond-mat. In an earlier decision
BCK-V-P-1262/02 of 24.03.2003, denoted latter as CK-03, CK has
formulated another reproach, No 2, that "the used by Radwanski
crystal-field approach is oversimplified, and the agreement of
calculations with experiments is accidental."

These two decisions have been undertaken by means of opinions
with the negative conclusion of Prof. Prof. H. Szymczak (November
2001), J. Sznajd (February 2003), Prof. A. M. Oles (December 2003)
and of J. Klamut (April 2004). For the final decision two last
opinions have been crucial. All of the referees belong to the
best polish solid-state physicists and magneticians, so such
opinions deserve on the serious attention by the magnetic
community, not only polish but the international one. Opinions
with the positive conclusion of Prof.  Prof. K. Krop (October
2001), R. Micnas (April 2002) and K. Wysokinski (December 2002)
have been found insufficient, in light of four negative opinions,
to provide evidence for substantial scientific achievements
required by the Polish law. \vspace {-0.3cm}
\section{Crystal field in modern solid-state physics and its extension to Quantum Atomistic Solid State Theory}
\vspace {-0.3cm} These decisions and objections become a part of
the long-lasting world discussion going on about the use and the
applicability of the CEF approach. This discussion lasts already
75 years as the CEF theory has been started in 1929 by Bethe and
followed by Kramers, Van Vleck and many, many others. Despite of
75 years and in meantime (1936-1938) formulation of the
competitive band theory there is no consensus within the modern
solid-state physics on the description of compounds containing
transition-metal atoms with open 3d/4f/5f shells. These compounds
exhibit so exciting phenomena like heavy-fermion behaviour at low
temperatures and unexpected, in frame of band models, insulating
ground state of 3d monoxides. In such scientific circumstances
the decision of CK disqualifying the crystal-field theory is,
according to us, premature and simply harmful to Physics. The
most important is a fact, that important polish physicists by
means of CK like to solve a serious scientific problem by means
of the administration decision. From such a point of view this
decision is a curious one as the European civilization already
370 years ago has learnt that no administrative inquisition-like
decision, even of the highest level, can solve any scientific
problem. We add that nobody has questioned in a scientific way
anyone of our published papers!!! We admit that we suffer often
an unscientific treatment of our submissions by Editors, who often
find them simply not suitable without a clear scientific
formulation of objections. We are sure, however, that a good
science will always win, i.e. will show its physical adequacy and
the conceptual fertility, and we also know from the history of
science that good theories suffer often seriously for a pretty
long time. Thus we continue our studies by more than 20 years and
we are optimists. As violation of scientific rules we presume the
rejection to publish a Comment, that corrects a recently published
paper. In a consequence, for instance, oversimplified electronic
structures of 3d ions, without strong correlations and without
the spin-orbit coupling, still appear in Phys. Rev. Lett. and
Phys. Rev. B despite of our (not suitable) submission
"Relativistic effects in the electronic structure for 3d
paramagnetic ions" PRL-LS 6925 from 1997 (available at ArXiv
cond-mat/9907140).

Theoretical hypothesis of our 20 years research can be formulated
as: macroscopic properties of compounds containing open-shell
3d/4f/5f atoms are predominantly determined by the low-energy
discrete electronic structure, with separations below 1 meV.
These states originate from atomic-like energy states of 3d/4f/5f
ions. For description of these atomic-like states the local
surroundings, crystal-field, spin-orbit and strong intra-atomic
correlations have to be taken into account.  As for description
of a crystalline solid it is necessary to consider at least the
translational symmetry and inter-site interactions, in particular
spin-dependent interactions indispensable for formation of
magnetically-ordered state, we have extended the CEF theory to an
Quantum Atomistic Solid State Theory (QUASST) \cite{4,5}. The CEF
theory is, however, the basic ingredient of QUASST, particularly
important for the physical understanding and the overall
scientific paradigm. Coming out with QUASST we would like to skip
somehow the crystal-field theory that has got a negative meaning
in the solid-state physics, becoming a synonym of the
oversimplified point-charge model. We hope that magnetic
theoreticians give some credit for QUASST to allow showing its
applicability and usefulness for understanding of
transition-metal compounds.

The crystal field gives explanation for the physical origin of the
observed low-energy electronic structure, yielding their nature
and a well-defined number. Strong intra-atomic correlations
assure that these states are describable for an atom being the
full part of  a solid like in the free ion. It means that in
QUASST we assume that the atomic-like integrity is preserved even
when the given atom becomes the full part of a solid. It is a
very strong assumption but taking into account that it is based
on the generally-accepted concept of the atomistic construction
of matter surely is worth to be thoroughly studied. We are
consequently doing it in the Center of Solid State Physics in
Krakow.  The valency of the atom in a solid depends on the
partner(s) and the stoichiometry. It can be 3+ in case of
Pr$_2$O$_3$, but 4+ for PrO$_2$. In metallic PrNi$_5$, without
judging the formal stoichiometry of Pr and Ni atoms, the observed
discrete electronic structure turns out to be related to the
4f$^2$ configuration occurring formally in the Pr$^{3+}$ ion.
\vspace {-0.3cm}
\section{The CEF approach and QUASST in conventional 4$\textrm{f}$}
\vspace {-0.3cm} Among others Radwanski and Franse in years
1984-1995 has put a substantial contribution to show the physical
adequacy of the CEF approach to conventional 4f compounds, like
Ho$_2$Co$_{17}$, Dy$_2$Co$_{17}$, Nd$_2$Fe$_{14}$B, ErNi$_5$,
DyNi$_5$, NdNi$_5$, PrNi$_5$, .. All of them are metallic. All are
magnetic, apart of PrNi$_5$ down to 1 K. By physical adequacy we
understand a highly consistent description of physical
properties. Let focus on (anisotropic) magnetic properties of all
above mentioned compounds. For it we correlated macroscopic
properties, like value of the magnetic moment and its direction
in the crystal, with atomic-scale properties like localized
states with (low-)energies and eigenfunctions. By it we could
prove that the observed huge anisotropy is predominantly of the
single-ion origin. The derived CEF-like electronic structure from
high-field magnetization measurements have been later positively
verified by specific heat measurements \cite{6}. A conical
structure of Nd$_2$Fe$_{14}$B below 140 K has been nicely
described within the CEF theory revealing the importance of
higher-order CEF interactions \cite{7}. The importance of
higher-order CEF interactions is manifest again in the
first-order metamagnetic transition at 17 T.

The model analysis of the overall temperature dependence of the
specific heat is shown in Fig. 1. Fig. 2 shows the fine electronic
structure of the 4f$^{11}$ configuration (the Er$^{3+}$ ion)
associated with the $^4$I$_{15/2}$ ground multiplet. The perfect
description both in magnetic and paramagnetic state with the
$\lambda$ peak at T$_c$ should be noted. Concluding ErNi$_5$ we
say that in metallic magnetic compound coexist localized electrons
having discrete states with conduction electrons originating from
outer shells of Er and Ni. Magnetic and electronic properties are
predominantly governed by localized electrons with states
determined by CEF interactions.
\begin{figure}[ht]
\begin{center}
\includegraphics[width = 9.5 cm]{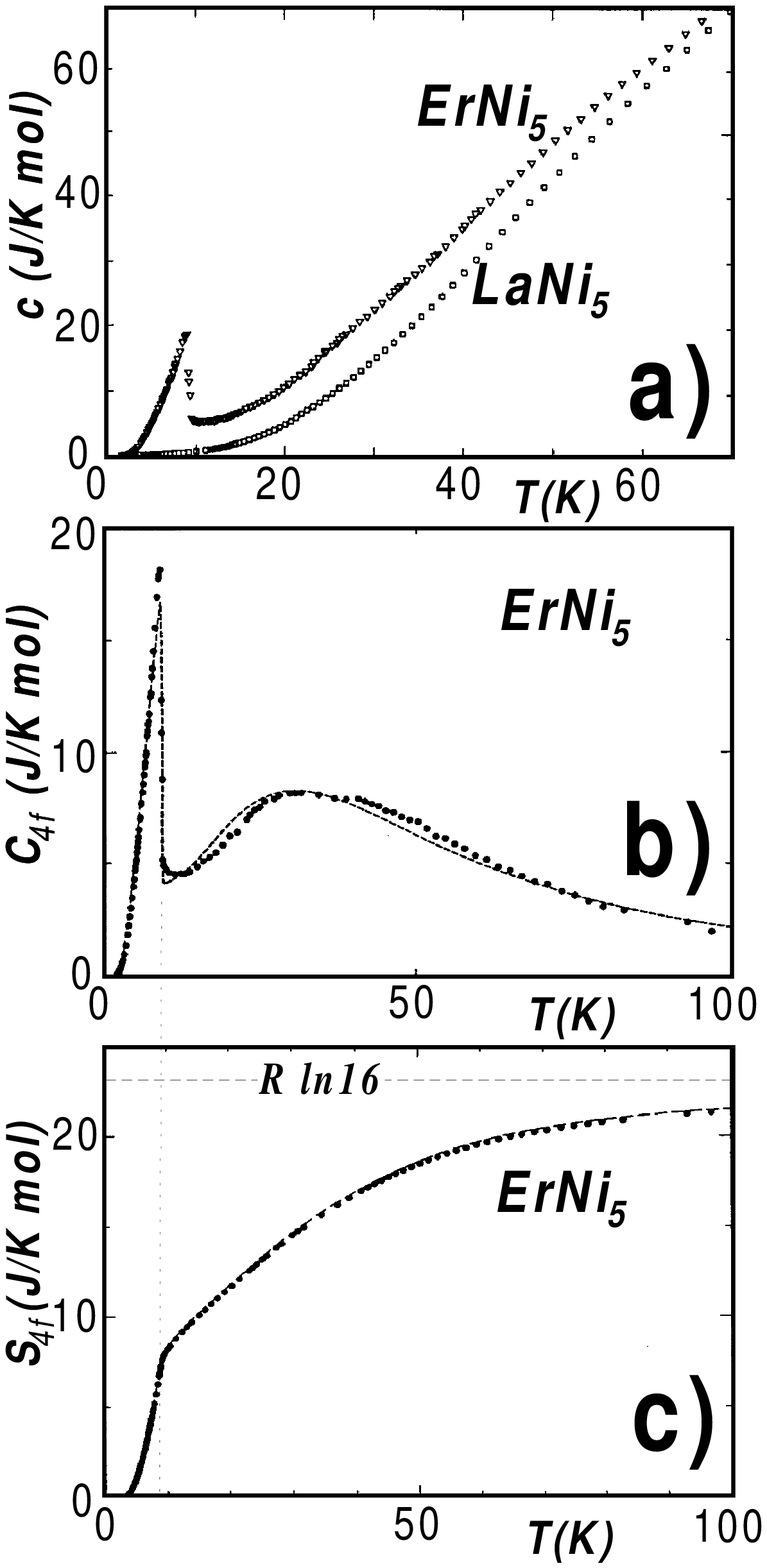}
\end{center}
\vspace {-1.5cm}\caption{Magnetic phase transition in ErNi$_{5}$.
a) Temperature dependence of the experimental heat capacity of
single crystalline ErNi$_{5}$ and LaNi$ _{5}$; b) Temperature
variation of the contribution of the $f$ subsystem to the heat
capacity of ErNi$_{5}$ (points - experimental data \cite{6}). The
dotted line shows the $f$-subsystem contribution calculated for
the atomic-like discrete energy spectrum determined by the strong
spin-orbit coupling, CEF and spin-spin interactions \cite{8}. }
\end{figure}

\begin{figure}[ht]
\begin{center}
\includegraphics[width = 8.2 cm]{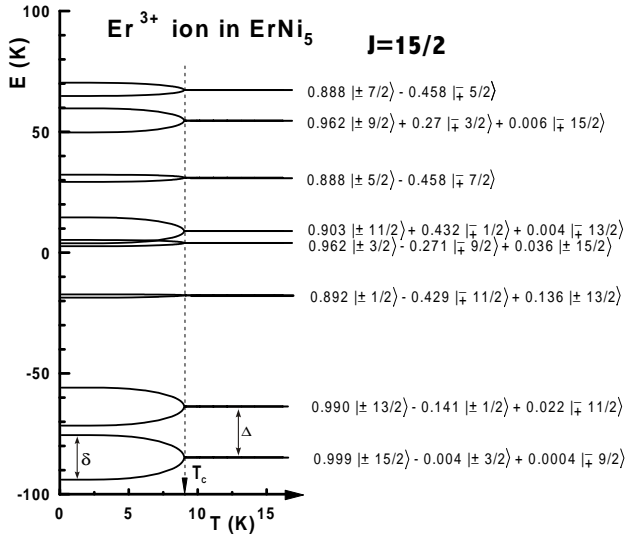}
\end{center}
\vspace {-0.5cm}\caption{The calculated fine electronic structure
of the 4f$^{11}$ configuration (the Er$^{3+}$ ion) in ErNi$_5$.
Below T$_c$ of 9 K the magnetic state is formed what becomes
visible by the splitting of Kramers doublets \cite{8}.}
\end{figure}
The CEF theory at the start points two things. One, that a solid
is not a homogeneous jellium (magma) but there exists varied in
space the electrostatic potential obviously due to charge
polarized atoms and electrons (the simplest version is a charge
point ionic model). Secondly, a 3d/4f/5f paramagnetic atom serves
as an atomic-scale agent to probe this potential. The CEF theory
points out the multipolar character of this electrostatic
potential. It is reflected in subsequent orders of CEF parameters
(quadrupolar - B$_2$$^0$, B$_2$$^2$ parameters; octupolar -
B$_4$$^0$, B$_4$$^4$, …, dodehexapolar B$_6$$^0$, B$_6$$^6$,
..).  For a paramagnetic ion this multipolar potential causes the
splitting of its ionic electronic structure. This splitting is a
hallmark of the CEF theory. This splitting in case of 4f
compounds is surprisingly well describable making use of the
total angular momentum quantum number J as the good quantum
number. Actually, we should work with the all-term electronic
structure instead of the one, Hund's rule, ground multiplet only.
The successful approximation with only one multiplet is due to
the strong spin-orbit coupling that causes the excited multiplet
to lie at least 0.3 eV above the ground multiplet preventing its
substantial thermal population at, say, room temperature.
Energies of this electronic structure can be later verified by,
for instance, specific heat measurements and by spectroscopic
measurements using inelastic neutron scatterings. The
eigenfunction of the ground state bears information about the
magnetic moment, its value and the direction, a fact that we
strongly employ in our studies.

As a strong confirmation of the CEF approach we take the
possibility of prediction of magnetic properties of the
isostructural compound with another rare-earth atom. Using the
single-ion scaling we had predicted in 1986, for instance, a
value of the transition field of 26 T for Dy$_2$Co$_{17}$ basing
on 19 T for Ho$_2$Co$_{17}$. In years 1991-1995 a remarkably
consistent description within the CEF approach has been obtained
for the RNi$_5$ series, both zero-temperature properties and
thermodynamics. All of the above mentioned compounds, except
PrNi$_5$, are magnetically ordered. Thus, the calling the applied
approach as the CEF approach is only a nick-name pointing out the
fundamental role of CEF states for the magnetic and electronic
properties. Of course, a magnetic order cannot be obtained within
the purely CEF approach. However, we know what happens to CEF
states when the magnetic order is formed. The magnetic state
develops on the CEF states. In fact, all of the analysis of
RNi$_5$ compounds illustrate the action of the QUASST theory.

Sub-Conclusion: there is wide experimental evidence for the
existence of CEF states in rare-earth (4f) compounds, both
metallic and ionic. From our studies of ionic compounds we can
mention Nd$_2$CuO$_4$ and ErBa$_2$CuO$_7$. \vspace {-0.3cm}
\section{Extension of the CEF approach to actinides (5$\textrm{f}$
compounds)} \vspace {-0.3cm} Just after the first experimental
results on newly discovered in group of Prof. Frank Steglich
heavy-fermion metal UPd$_2$Al$_3$ Radwanski and Franse  in 1992
have described the specific heat, from 4 to 300 K,  as related to
the 5f$^3$ (U$^{3+}$) configuration. We have managed to describe
the overall temperature dependence with a Schoottky-like peak at
50 K and a $\lambda$-type peak related to the antiferromagnetic
state formed at T$_N$ of 14 K. This energy level scheme has been
confirmed by INS experiment of Krimmel/Steglich in 1996 \cite{9}
as we pointed out in year of 2000 \cite{10}. The observation of
well-defined localized CEF excitations in heavy-fermion metal
UPd$_2$Al$_3$ we take as great confirmation of our atomistic
approach. With great pleasure we have noted in year of 2001 a
change of mind of Fulde and Zwicknagl from the itinerant picture
for all f electrons to a dual model with two fully localized f
electrons \cite{11}. Two or three localized electrons we treat as
a minor problem, because the main theoretical difference is
related to the itinerant or localized point of view. For the
scientific honesty we have to mention that the problem of
localized states in UPd$_2$Al$_3$ is not yet over - another
German group of Lander with coworkers quite recently claim that
there is no evidence for the localized states in UPd$_2$Al$_3$
\cite{12}. Just after appearance of these doubts we again clearly
defined our point of view and our interpretation with the 5f$^3$
configuration \cite{13}.

\begin{figure}[ht]
\begin{center}
\includegraphics[width = 8.7 cm]{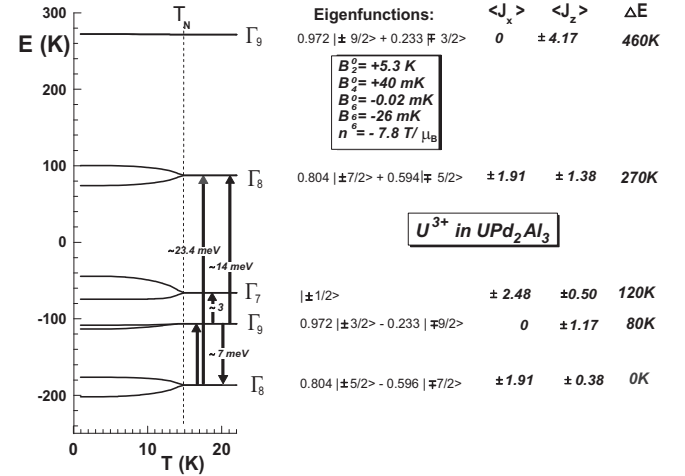}
\end{center}
\vspace {-0.5cm}\caption{Energy level scheme of the 5f$^3$ (the
U$^{3+}$ ion) in UPd$_2$Al$_3$ taken from Refs \cite{10,13}.
Arrows indicate transitions which we have attributed to
excitations revealed by inelastic-neutron-scattering experiments
of Krimmel {\it et al.} \cite{9}.}
\end{figure}

We prefer 3 f electrons owing to the intrinsic dynamics of the
Kramers system, states of which are established by the atomic
physics (in particular the number of states and their
many-electron atomic-like nature). We can add that no one
succeeded in description of the observed transitions and other
properties to the 5f$^2$ configuration with the similar
consistency to ours.

The derived electronic structure accounts, apart of the INS
excitations, also surprisingly well for the overall temperature
dependence of the heat capacity, the substantial uranium magnetic
moment and its direction. We make use of a single-ion like
Hamiltonian, the same as has been used for ErNi$_5$, for the
ground multiplet J=9/2 \cite{6,16}:
\begin{center}
$H=H_{CF}+H_{f-f}=\sum \sum B_n^mO_n^m+n_{RR}g^2\mu _B^2\left(
-J\left\langle J\right\rangle +\frac 12\left\langle J\right\rangle
^2\right) $
\end{center}
\vspace {-0.2cm} The first term is the crystal-field Hamiltonian.
The second term takes into account intersite spin-dependent
interactions (n$_{RR}$ - molecular field coefficient, $g$=11/8 -
Lande factor) that produce the magnetic order below T$_N$ what is
seen in Fig. 3 as the appearance of the splitting of the Kramers
doublets and in experiment as the $\lambda$-peak in the heat
capacity at T$_{N}$.

The splitting energy between two conjugate Kramers ground state
agrees surprisingly well, both the value of the energy and its
temperature dependence, to a low-energy excitation of 1.7 meV at
T=0 K observed by Sato {\it et al.} \cite{14,15}, which has been
attributed by them to a magnetic exciton. Thus, we are convinced
that the 5f$^{3} $(U$^{3+}$) scheme provides a clear physical
explanation for the 1.7 meV excitation (magnetic exciton) - this
excitation is associated to the removal of the Kramers-doublet
ground state degeneracy in the antiferromagnetic state.

For actinides we also should mention the consistent description
(5f$^3$, U$^{3+}$) of a ferromagnetic metal UGa$_2$, T$_c$ of 125
K, both zero-temperature properties (magnetic moment of 2.7
$\mu_B$ lying in the hexagonal plane, T$_c$=125 K) and
thermodynamics (temperature dependence of the specific heat and
of anisotropic paramagnetic susceptibility) \cite{16}. Later this
description has been extended to NpGa$_2$ (5f$^4$, Np$^{3+}$),
isostructural easy axis ferromagnet properties of which has been
described using the single-ion correlation (Stevens factors)
\cite{17}.

Recently in autumn of 2003 a well-defined localized excitation
has been observed in heavy-fermion metal YbRh$_2$Si$_2$.
Sichelschmidt from F. Steglich group has managed to observe in
this heavy-fermion metal at temperature T=1.5 K an ESR signal
typical for the localized Yb$^{3+}$ ion \cite{18}. The observation
of the ESR signal is a large surprise as YbRh$_2$Si$_2$ was
regarded as a prominent heavy-fermion metal with the Kondo
temperature T$_K$, of 25-30 K. The Kondo model does not expect
localized states to exist at temperatures lower than T$_K$,
whereas temperature of 1.5 K is more than 10 times smaller than
T$_K$. Surely, such the observation calls for the rejection, or
at least a substantial revision of the Kondo lattice theory. We
are convinced that this revision will go to our CEF based
understanding of the heavy-fermion phenomena with the importance
of the local Kramers doublet ground state. Just after the
Sichelschmidt/Steglich discovery we have described the g tensor
and derived two sets of CEF parameters for $\Gamma$$_6$ and
$\Gamma$$_7$ CEF ground states \cite {19}.

Sub-conclusion: We take these examples as further evidence for
the applicability of the CEF-based approach to actinides and
anomalous 4f/5f compounds. Our basic idea for the localized CEF
origin of heavy-fermion phenomena has been formulated already in
1992 \cite{20}. A report of CSSP-4/95 "Physics of heavy-fermion
phenomena" \cite{21} has been widely distributed to the leading
scientists over 400 copies, including the International Board of
SCES-94 and SCES-95. Our CEF-based interpretation of the
heavy-fermion phenomena has been put in 1995 to the scientific
protection of the Prezes of the Polish Academy of Science.
\vspace {-0.5cm}
\section{Anomalous properties and heavy-fermion behavior}
\vspace {-0.3cm} In our understanding of anomalous 4f/5f compounds
the localized Kramers doublet ground state plays the essential
role \cite{20,21}. A lattice of Kramers ions with the local
Kramers doublet ground state is the physical realization of the
anisotropic spin liquid postulated {\it ad hoc} in heavy-fermion
theories. According to us, the heavy-fermion behavior is related
to difficulties in the removal of the Kramers doublet degeneracy.
The local Kramers doublet is always formed for a
strongly-correlated odd-number electron system. The removal of
the Kramers degeneracy is equivalent to the formation of the
magnetic state, characterized by breaking of the time-reversal
symmetry. There can be different reason for this difficulty in
the removal of the Kramers degeneracy (this difficulty can be
called as a quantum entanglement of two Kramers conjugate states)
causing its removal at low temperatures only. The Kramers-doublet
degeneracy has to be removed before the system approaches zero
temperature. In this view heavy-fermion state is a magnetic
state. In contrast to well-defined magnetic/paramagnetic
transition characterized by the lambda-type peak in the specific
heat the magnetic state in heavy-fermion compounds is not
uniformed, being of the spin-fluctuation type. There is a
site-to-site change of value of the Kramers doublet splitting.
Associated with it is a site-to-site change of the value of the
local magnetic moment and its direction. Thus one can model such
magnetic state by a statistical distribution of the 0-0.3 meV
splittings and of Kondo temperatures. In QUASST heavy-fermion
excitations are neutral spin-like excitations between conjugate
local Kramers states. These thermal excitations are associated
with the reversal of spin. In our picture f electrons (exactly f
electron states) are localized , whereas f excitations looks like
itinerant (no one can say which exactly atom becomes excited).
Our explanation with localized f electrons is unpopular within
the magnetic community which prefers itinerant f electrons. If f
electrons would be really itinerant than the conductivity of a
heavy-fermion compound would be larger than the reference La/Y/Lu
compound. But in experiment is always opposite - the resistivity
of a heavy-fermion compound is always larger than the reference
system. Finally we can add that in QUASST CEF-like f states do
not lie at the Fermi energy. The Fermi surface is established for
itinerant conduction electrons only. Moreover, in QUASST the
heavy-fermion like phenomena at low temperatures can occur also
in ionic compounds. This analysis of anomalous and heavy-fermion
behavior in transition-metal compounds was not a subject of
evaluation by CK - here it was added for the completeness
reasons. \vspace {-0.3cm}
\section{3$\textrm{d}$ ionic compounds}
\vspace {-0.3cm} In 1996 we have realized that a standard approach
to electronic structures and magnetism of 3d-ion compounds
substantially differ from that used in rare-earth compounds.
Namely, owing to the weakness of the spin-orbit coupling, in 3d
compounds the spin-orbit coupling has been customarily ignored.
As a consequence the magnetic moment was essentially of the
spin-type only whereas the electronic structure was built from
the orbital-only states. Moreover, the concept of electronic
structures and magnetism was built on single-electron states,
t$_{2g}$ and eg orbitals known from the octahedral crystal field,
with neglecting intra-atomic electron correlations among d shell.
In 1997 we have performed calculations for the spin-orbit effect
on the electronic states of 3d paramagnetic ions in the
octahedral crystal field revealing a variety of low-energy
states, Fig. 4 \cite{22,23}. For these calculations we have taken
into account strong correlations among electrons in the 3d shell
by considering many-electron states and two Hund's rules. These
structures have been put in 2000 to the scientific protection of
the President of the American Physical Society.

According to the Quantum Atomistic Solid-State theory the
atomic-like electronic structures, shown in (c), are preserved
also in a solid. The shown states are many electron states of the
whole d$^n$ configuration. At zero temperature only the lowest
state is occupied. The higher states become populated with the
increasing temperature. In Fig. 4 on the lowest levels the
magnetic moment (in $\mu _{B}$~) are written. Their are not
integer. It means that a general conviction that the localized
model gives the magnetic moment of the unpaired n localized d
electrons as 2n$\mu _{B}$ (or (10-2n)$\mu _{B}$) is not true.
\begin{figure}[ht]
\begin{center}
\includegraphics[width = 8.2 cm]{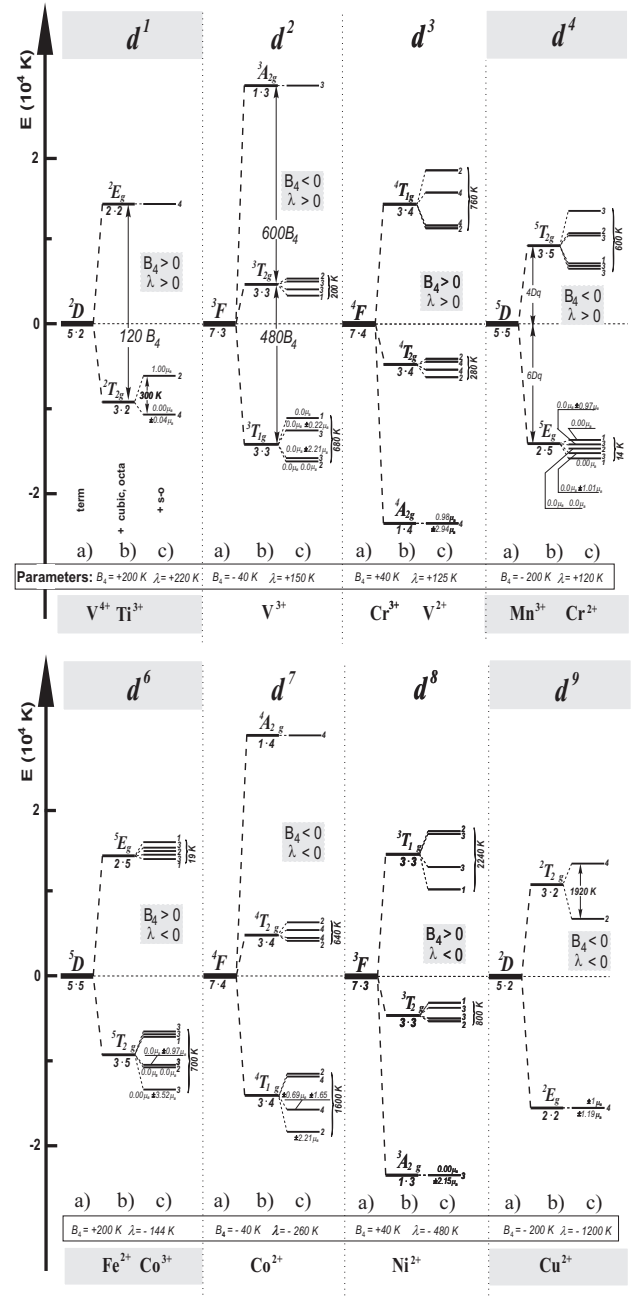}
\end{center}
\vspace {-0.5cm}\caption{The calculated electronic structure of
the 3d$^{n}$ configurations of the 3d ions, 1$\leq n$ $\leq 9$, in
the octahedral crystal field (b) and in the presence of the
spin-orbit coupling (c). (a) - shows the Hund's rule ground term.
Levels in (c) are labeled with degeneracies in the LS space
whereas in (b) the degeneracy is shown by the orbital spin
degeneracy multiplication. The spin-orbit splittings are drawn
not in the energy scale that is relevant to CEF levels shown in
figures b \cite{22,23}.}
\end{figure}

This approach called a strongly-correlated crystal field approach
[23] is in contrast to the single-electron crystal-field approach
customarily presently used. By doing it we have made unification
of 3d and rare-earth compounds in description of the low-energy
electronic structures and magnetism, of course keeping the
relevant strength of the spin-orbit coupling. We have calculated
the low-energy electronic structure and correlate it with
magnetic and electronic properties, e.g. $^{3}T$$_{1g}$ of the
V$^{3+}$ ion in LaVO$_3$ and $^{5}$E$_g$ of the Mn$^{3+}$ ion in
LaMnO$_3$. In SCES-02 there was a reproach to us that these
ground subterms are incorrect owing to literature t$^2$$_{2g}$
and t$^3$$_{2g}$e$_{g}$ configuration, with the e$_{g}$ orbital
higher whereas derived by us $^{5}$E$_g$ subterm is lower. Despite
of our long explanations, explaining lower and capital symbols,
the papers have been rejected - the International Advisory Board
have been informed about this controversy by the Chairman of
SCES-02. Just after, our solution with the ground subterm
$^{3}T$$_{1g}$ for the V$^{3+}$ ion (3d$^2$ configuration) in
LaVO$_3$ or V$_2$O$_3$ and $^{5}E$$_{g}$ for the Mn$^{3+}$ ion
(3d$^4$ configuration) in LaMnO$_3$ has been put to the scientific
protection of the Rector of the Jagiellonian University in Krakow
and of the AGH University of Mining and Metallurgy. Recently A.
M. Oles, the vice chairman of SCES-02, has admitted the
correctness of our ground states in LaVO$_3$ (V$_2$O$_3$) and in
LaMnO$_3$. We await for further scientific steps.

We have clarified the electronic structure and magnetism of
LaCoO$_3$ \cite{24}. It turns out that relatively strong
octahedral crystal field yields the breaking of Hund's rules
establishing the ground subterm $^{1}$A$_1$ originating from
$^1$I term that in the free Co$^{3+}$ ion lies 4.45 eV above the
ground term. However, the octahedral crystal field interactions
are too weak to break intra-atomic correlations and to create
conditions for the applicability of the single-electron approach.
It means that the experimentally observed states are still
described within the atomistic QUASST approach. By perfect
reproduction of the ESR results of Noguchi from 2002 \cite{25} on
the excited triplet we have proved that this triplet is a part of
the $^5$T$_{2g}$ sub-term, originating from the high-spin $^5$D
term. It means that there is no intermediate spin-state, with S=1,
despite of theoretical LDA-U calculations of Korotin et al. from
1996 \cite{26} and a numerous literature on this subject. Thus, we
have confirmed the substantial physical applicability of the
atomistic CEF-based Tanabe-Sugano diagrams, existing already 50
years, applicability of which have been questioned by
band-structure calculations. The breaking of the Hund's rules in
LaCoO$_3$ results from the extraordinary small Co-O distance, of
192 pm. In CoO, for instance, the Co-O distance amounts to 217 pm.
This smaller distance by 13$\%$ causes increase of B$_4$$^0$
parameter by 85$\%$ owing to the R$^{-5}$ dependence of the
octupolar CEF interactions.

We have calculated the orbital moment in NiO \cite{27}, CoO
\cite{28}, FeBr$_2$ and LaMnO$_3$\cite{29}. We have derived highly
anisotropic properties of these compounds in full agreement with
the experimental evidence.
\begin{figure}[ht]
\begin{center}
\includegraphics[width = 6.2 cm]{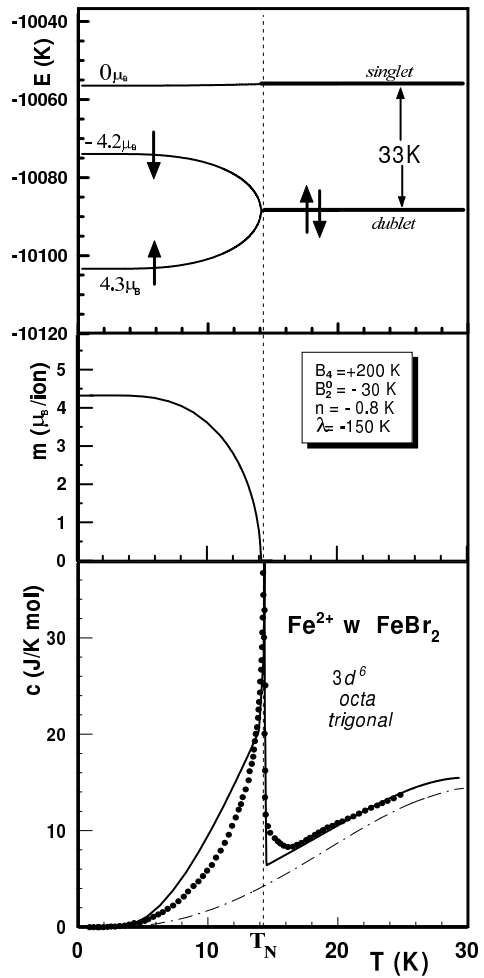}
\end{center}
\vspace {-0.5cm}\caption{Magnetic phase transition in FeBr$_{2}$.
a) temperature dependence of the 3 lowest energy levels
originating from the $^5$T$_{2g}$ subterm of the Fe$^{2+}$ ion; b)
temperature dependence of the magnetic moment of the Fe$^{2+}$
ion; c) temperature variation of the contribution of the $d$
subsystem to the molar heat capacity. The points represent the
experimental data \cite{30}.}
\end{figure}
In Fig. 5 we present the calculated temperature dependence of the
specific heat of FeBr$_2$ both in the paramagnetic and
antiferromagnetic state with the $\lambda$- peak at T$_N$ of 14 K
\cite{30}. For description we use the similar Hamiltonian as
shown above. The formation of the magnetic state is related with
the splitting of the lowest quasi triplet. It is worth to add that
this quasi-triplet is the excited quasi-triplet discussed above
for LaCoO$_3$. The Fe$^{2+}$ and Co$^{3+}$ ions are isoelectronic
3d$^6$ systems. Despite of the hexagonal lattice symmetry of
FeBr$_2$ and the slightly distorted cubic structure of LaCoO$_3$
in both these compounds the local symmetry is octahedral. We
would like to turn attention that the good reproduction of the
overall specific heat means, in fact, the counting of atoms. The
reproduction of the absolute value indicate that all atoms
equally contribute to the observed property. It means, though it
could sound unbelievedly, that all Fe atoms have the same
electronic structure. We think that it is an effect of the blind
action of the simple physical laws.

Recently we describe consistently NiO within the
strongly-correlated CEF approach reconciling its insulating
ground state, the value and the direction of the magnetic moment
in the antiferromagnetic state below 525 K and thermodynamic
properties \cite{31}. In particular we have calculated the
overall electronic specific heat with the lambda-type peak at
T$_N$ and a substantial heat with the overall entropy of Rln3
fully released at T$_N$. We have quantified crystal-field (the
leading parameter B$_{4}$ = +21 K), spin-orbit (-480 K, i.e. like
in the free ion \cite{32}) and magnetic interactions (B$_{mol}$
of 503 T and n= -200 T/$\mu_{B}$). In our approach E$_{dd}$ $\gg$
E$_{CF}$(=2.0 eV)$\gg$E$_{s-o}$(=0.29 eV)$\gg$E$_{mag}$(=0.07
eV). The orbital and spin moment of the Ni$^{2+}$ ion in NiO has
been calculated within the quasi-atomic approach. The orbital
moment of 0.54 $\mu _{B}$ amounts at 0 K in the
magnetically-ordered state, to about 20\% of the total moment
(2.53 $\mu _{B}$). Despite of using the full atomic orbital
quantum number $L$=3 and $S$=1, the calculated effective moment
from the temperature dependence of the susceptibility amounts to
3.5-3.8 $\mu _{B}$, i.e. only 20 $\%$ larger value than a
spin-only value of 2.83 $\mu _{B}$.

We take as great confirmation of the strongly-correlated crystal
field approach that the electronic structure through the series
of compounds results from the symmetry of the transition-metal
surroundings. For instance, in all of above mentioned compounds
the closest surroundings has predominantly the octahedral
symmetry. As a consequence the ground state subterm alternates as
the octupolar moment of the 3d$^n$ configuration.

\vspace {-0.5cm}
\section{Development of physics}
Everybody can see that our understanding of the electronic
structure and magnetism of transition-metal compounds is along
the well-established in solid-state physics paradigm. By pointing
out the importance of the crystal field we call for a larger
attention to local atomic-scale effects. In the present situation
of the administrative reproaching of the use and the concept of
the crystal-field approach if applied to magnetic metallic
compounds we call to members of the international physical
community. We have to turn to the international community,
because by years Polish physicists can not settle up a problem
and by years after the first negative Szymczak's opinion of 2001
subsequent opinions are written in very loosely language. For
instance in the opinion of A.M. Oles, crucial for the final
negative decision, there is only one (!!) verifable statement,
that fitting of the only one physical property is not reliable.
We fully agree with this general Oles statement, but it does not
touch the scientific approach of Radwanski, where always we
analyze many properties for one compound and look for the
consistency with other isostructural compounds. Other reproaching
statements of Oles are not scientifically conclusive in the sense
that these statements contain always "seems to", "probably" or
"likely". Even a reproach of H. Szymczak, obviously erroneous,
about description of the trigonal off-octahedral distortion used
by Radwanski cannot be clarified. In our paper by Ropka and
Radwanski in Phys. Rev. B (63 (2001) 172404), Ref. 1 on the list
of our publication from 2001, we have described the trigonal
off-octahedral distortion, at the end of the first page, by the
B$_2$$^0$O$_2$$^0$ term added to the octahedral Hamiltonian
written for the z axis along the cube diagonal (reproach No 3).
This procedure has been simply erroneously questioned by Szymczak
in 2001, but there was no one to clarify it. According to normal
scientific rules Szymczak has been obliged to write a Comment to
the Editor of Phys. Rev. B and to inform publicly the scientific
community about incorrectness before writing an unsubstantiated
reproach to the Governmental Scientific Committee. (In fact, the
simplest was to send e-mail to Radwanski, and trying to explain
the controversy.) An unscientifically made reproach cannot be
clarified now. Later, four referees did not clarify this
erroneous reproach - we take it as an evidence for their
scientific dishonesty. At present instead of the simple
clarifying the erroneous decision of CK, the easiest would be the
correction by Szymczak, the Polish magnetic community keeps a
long-lasting splendid quiet. This unscientific behavior is partly
related to the dominant administrative position of H. Szymczak in
the Polish physics and a lack of respect for basic scientific
rules in Polish physics. We hope that these problems will be
discussed in the coming magnetic conferences in Poland, in
Wroclaw, 19-21 May 2005, on "Anomalous properties of strongly
correlated systems" chaired by D. Kaczorowski and in Poznan,
25-29 June 2005 on "Physics of Magnetism" chaired by Krompiewski
and R. Micnas.

We submit this problem also to Prof. E. Bauer, the Chairman of
the incoming SCES-05 Conference to be held in 26-30 July 2005 in
Vienna, and to all members of the International Advisory Board.
In particular, we turn to well-experienced physicists: M.
Abd-Elmeguid, P. Alekseev, J. Allen, M.C. Aronson, P. Coleman, M.
Continentino, B. Coqblin, A. de Visser, C. Di Castro, Z. Fisk, J.
Flouquet, A. Fujimori, P. Fulde, J. Gomez-Sal, H. Harima, H.
Johannesson, B. Johansson, C. Lacroix, A. Loidl, G. Lonzarich,
M.B. Maple, F. Marabelli, K.A. McEven, A.J. Millis, J.A. Mydosh,
Y. Onuki, G. Oomi, R. Osborn, M. Reiffers, T.M. Rice, T.F.
Rosenbaum, E.V. Sampathkumaran, H. Sato, G.A. Sawatzky, V.
Sechovsky, J. Sereni, M. Sigrist, J. Spalek, F. Steglich, T.
Takabatake, J.D. Thompson, K. Ueda, D. Vollhardt, H. v.
Lohneysen, V. Zlatic. All of us knows that the magnetism of
transition-metal compounds is still under scientific discussion
and any administration decision about the incorrectness of the
crystal-field-based approach to magnetic metallic materials is
premature and harmful to physics. A reproach that "the used
(crystal-field) approach is too simplified" with a simultaneous
statement that "the obtained agreements with experimental data
are accidental" is illogical. We are ashamed that such strong
administrative interference to Physics happens in Poland, a
country of the long tradition of freedom. We are lucky that apart
of the administration we have another great authority - the Pope
John Paul II. By last 20 years he teaches about the truth, the
values, the dignity and the freedom in everyday life and in
Science.

We turn to the international physical community as we believe
that all members of this community share our view that Science
can develop only in the truth and in freedom. We call to
physicists, our colleagues in searching for the scientific truth:
We can differ in approaching to Physics and Science - but all of
us agree that Science and Physics can develop only without
administration interference for judging correctness of any
scientific theory. We believe that the future proves the
incorrectness of the administrative interference to physics and
for the restoration of normal scientific conditions in physics.
Independently on it, we will continue with the highest integrity
our research on the magnetism and electronic structures of
transition-metal compounds being open for scientific discussions
and critics.

We are grateful to all opponents - thanks them our studies turn
out to be scientifically important despite of using at the start
well-known atomistic approach and the 75 years old crystal-field
theory. Being grateful to our opponents we cannot, however,
accept discrimination and inquisition methods used in doing
science. It is obvious that Science without ethic values becomes
empty. If somebody is able to prove errors in the crystal-field
approach is welcome to publish it openly. If somebody has
something against me and my scientific activity is welcome to say
it publicly, not to work with the help of administration methods
like rejection of papers from the publication or the presentation
on conferences. A great ethic problem appears if later he
publishes quite similar things. We declare our deep will to
cooperate with everybody in serving to built the scientific
magnetic community and to search for the scientific truth.

\section{Remarks on the crystal-field theory}
\vspace {-0.3cm} A strange scientific climate about the
crystal-field theory in the modern solid-state paradigm comes from
a widely-spread view within the magnetic community that it does
not have the proper theoretical justification. An oversimplified
point charge model is treated as an essence of the crystal-field
theory. An indication in some cases that the point charge model is
not sufficient to account for the crystal-field splittings was
taken as the conclusive proof for the incorrectness of the
crystal-field concept at all. According to us the theoretical
background for the crystal-field theory is the atomic construction
of matter. Simply, atoms constituting a solid preserve much of
their atomic properties. One can say that the atomic-like
integrity is preserved, after giving up partly or fully some
electrons, and then the atomic identity serves as the good quantum
number of the electron system. We are quite satisfied that the
point-charge model provides the proper variation the ground states
going on from one to another 3d/4f/5f atom. Different ionic states
we consider as different states of the atom, though it is better
instead of the ionic state to say about the electron
configurations and their different contributions to magnetic,
electronic, spectroscopic and optical properties. For instance, in
metallic ErNi$_{5}$ there exists 4f$^{11}$ electron configuration,
often written as the Er$^{3+}$ ion, that is found to be
predominantly responsible for the magnetism and the electronic
structure of the whole compound \cite{6}; the other electrons of
Er and Ni are responsible for the metallic behavior. We point out
the multipolar character of the electric potential in a solid. It
is very fortunate situation when a solid, with milliard of
milliard of atoms (in America billion of billions), can be
described with the single electronic structure. It is true, that
the crystal-field theory being itself a single-ion theory cannot
describe a solid with collective interactions. For this reason we
came out with the Quantum Atomistic Solid State Theory and
completed the crystal-field theory with strong intra-atomic
correlations and intersite spin-dependent interactions. By
pointing out the importance of the CEF theory we would like to
put attention to the fundamental importance of the atomic physics
(Hund's rules, spin-orbit coupling, ....) and local single-ion
effects. It is worth remind that the source of a collective
phenomenon, the magnetism of a solid, are atoms constituting this
solid. Properties of these potentially-active atoms (open-shell
atoms) are determined by local surroundings and local symmetry.
Subsequently, these atomic moments, with spin and orbital parts,
enter to the collective game in a solid. If somebody thinks that
CEF and QUASST is too simple (in fact, it is not simple!!) should
not blame authors for it, but Nature. Nature turns out to be
simpler than could be!

Another wrong conviction about the crystal-field theory is that it
was exploited already completely. In order to shown that this
thinking is wrong we turn the reader's attention that the
crystal-field approach used within the rare-earth and actinide
community (4f and 5f systems) fundamentally differs from that used
within the 3d community. The 4f/5f community works with $J$ as the
good quantum number whereas the 3d community "quenches" the
orbital moment and works with only the spin $S$. Our description
of a 3d-atom compound like FeBr$_{2}$ and LaCoO$_{3}$ one can find
in Refs \cite{30} and \cite{24}. In case of the
strongly-correlated crystal-field approach we work with
many-electron states of the whole 4f$^{n}$, 5f$^{n}$ 3d$^{n}$
configuration in contrary to single-electron states used in 3d
magnetism and LDA, LSDA, and many other so-called {\it ab initio}
approaches. Technically, strong correlations are put within the
CEF theory, and in QUASST, by application of two Hund's rules. The
{\it ab initio} calculations will meet the CEF (QUASST) theory in
the evaluation of the detailed charge distribution within the
unit cell and after taking into account strong intra-atomic
correlations among electrons of incomplete shells and the
spin-orbit coupling in order to reproduce the CEF conditions (two
Hund's rules, also the third one for rare-earths and actinides).

We would like to mention that we are fully aware that used by us
the Russell-Saunders LS coupling can show some shortages in case
of actinides related to the growing importance of the j-j
coupling. We are aware of many other physical problems which we
could not mention here due to the length problem - finally we
mention only that we can reverse scientific problem in the solid
state physics and use 4f/5f/3d compounds as a laboratory for the
atomic physics for study 3d/4f/5f atoms in extremal electric and
magnetic fields. In the solid-state physics we study the lowest
part of the atomic structure, but extremely exactly.

The detailed electronic structure is predominantly determined by
conventional interactions in a solid: the Stark-like effect by
the crystalline electric field potential due to 3-dimensional
array of charges in a crystal acting on the aspherical incomplete
shell, and the Zeeman-like effect due to spin-dependent
interactions of the incomplete-shell spin (atomic-like moment)
with self-consistently induced spin surroundings. These states
can become broaden in energy by different interactions (lowering
of the local symmetry, thermal expansion, appearance of a few
inequivalent sites, lattice imperfection, surface effects and
other solid-state effects). Obviously, we should not think that
discrete crystal field states mean that they are extremely thin
lines. 3 or even 10 meV broad lines are still of the
crystal-field origin. Underlying by us by many, many years the
importance of the crystal field we have treated as an opposite
view to the overwhelmed band structure view yielding the
spreading of the f-electron (and mostly 3d-electron) spectrum by
2-5 eV. \vspace {-0.4cm}

\section{Conclusions} \vspace {-0.3cm} We advocate for the high adequacy of use and the concept of the
crystal field approach, even if applied to metallic magnetic
materials of transition-metal compounds. By extension of the
crystal-field theory to a Quantum Atomistic Solid-State theory
(QUASST) we have made the unification of 3d and rare-earth
compounds in description of the low-energy electronic structures
and magnetism of open 3d/4f/5f shell electrons taking into
account the local crystal field, the intra-atomic spin-orbit
coupling and strong intra-atomic correlations. QUASST offers
consistent description of zero-temperature properties and
thermodynamic properties of 3d-ion containing compounds. We have
calculated the orbital moment in 3d oxides (NiO, CoO, LaCoO$_3$,
LaMnO$_3$, FeBr$_2$). Our studies indicate that it is the highest
time to unquench the orbital magnetism in 3d -ion compounds. We
claim that the first-principles and ab initio studies will be as
long not successful as strong correlations assumed in the CEF
approach will be not incorporated in the calculations. We do not
claim to (re-)invent crystal-field theory, but we do not agree
for the depreciation of the crystal field theory. We are
convinced that the crystal-field theory with strong correlations
is a fundamental ingredient of the modern solid-state paradigm.

$^\spadesuit$ dedicated to the Pope John Paul II, a man of
freedom in life and in Science.


\begin{thebibliography}{9}
\bibitem{1} P. Mohn, {\it Magnetism in the Solid State}, Solid State Sciences \textbf{134} (Springer-Verlag, Berlin) 2003.

\bibitem{2} J. H. Van Vleck, {\it Theory of electric and magnetic susceptibilities} (Oxford University Press)
1932.

\bibitem{3} Y. Tanabe and S. Sugano, J. Phys. Soc. Japan \textbf{9},
753 (1954); S. Sugano, Y. Tanabe, and H. Kamimura, {\it
Multiplets of Transition-Metal Ions in Crystals} (Academic, New
York) 1970.

\bibitem{4} R. J. Radwanski and Z. Ropka, {\it Quantum Atomistic Solid State Theory},
cond-mat/0010081.

\bibitem{5} R. J. Radwanski, R. Michalski, and Z. Ropka, Acta Phys. Pol. B
\textbf{31}, 3079 (2000).

\bibitem{6} R. J. Radwanski, N. H. Kim-Ngan, F. E. Kayzel, J. J. M. Franse, D. Gignoux, D. Schmitt,
F. Y. Zhang, J. Phys.: Condens. Matter \textbf{4}, 8853 (1992).

\bibitem{7} R. J. Radwanski and J.J.M. Franse, J.Magn. Magn. Mat. \textbf{80}, 14 (1989).

\bibitem{8} R. J. Radwanski, R. Michalski, Z. Ropka, and A. Blaut, Physica B \textbf{319}, 78 (2002).

\bibitem{9} A. Krimmel, A. Loidl, R. Eccleston, C. Geibel and F. Steglich,
J. Phys.: Condens. Matter \textbf{8}, 1677 (1996).

\bibitem{10} R. J. Radwanski, R. Michalski and Z. Ropka, Physica B
\textbf{276-278}, 803 (2000); R. J. Radwanski and Z. Ropka,
Czech. J. Phys. \textbf{54}, D295 (2004); cond-mat/0407453.

\bibitem{11} G. Zwicknagl, A. Yaresko and P. Fulde, Phys. Rev. B \textbf{68} (2003) 052508;
Phys. Rev. B \textbf{65}, 081103(R) (2002); G. Zwicknagl and P.
Fulde, J. Phys.: Condens. Matter \textbf{15}, S1911 (2003).

\bibitem{12} A. Hiess, N. Bernhoeft, N. Metoki, G.H. Lander, B. Roessli, N.K.
Sato, N. Aso, Y. Haga, Y. Koike, T. Komatsubara, and Y. Onuki,
cond-mat/0411041; N. Bernhoeft, A. Hiess, N. Metoki, G.H. Lander
and B. Roessli, cond-mat/0411042.

\bibitem{13} R. J. Radwanski and Z. Ropka, {\it Are there crystal field levels in UPd$_2$Al$_3$?} {\it We answer, THERE
ARE}, cond-mat/0412257.

\bibitem{14} N. K. Sato, N. Aso, K. Miyako, R. Shiina, P. Thalmeier, G.
Vareloglannis, C. Geibel, F. Steglich, P. Fulde, T. Komatsubara,
Nature \textbf{410}, 340 (2001).

\bibitem{15} F. Steglich, N. K. Sato, N. Aso, P. Gegenwart, J. Custers, K. Neumaier, H. Wilhelm,
C. Geibel, O. Trovarelli, Physica B \textbf{329-333}, 441 (2003).

\bibitem{16} R. J. Radwanski and N. H. Kim-Ngan, J. Alloys-Comp. \textbf{219}, 260 (1995).

\bibitem{17} R.J. Radwanski, J.Phys.:Condens. Matt. \textbf{8},
10467 (1996).

\bibitem{18} J. Sichelschmidt, V.A. Ivanishin, J. Ferstl, C.
Geibel, and F. Steglich, Phys. Rev. Lett. \textbf{91}, 156401
(2003).

\bibitem{19} R. J. Radwanski and Z. Ropka, {\it Crystal-field ground state for the localized f
state in heavy fermion metal YbRh$_2$Si$_2$}, cond-mat/0312725.

\bibitem{20} R. J. Radwanski, {\it Formation of the heavy-fermion state - an explanation in
a model traditionally called localized} - presented at SCES-92 in
Sendai as 8P92; cond-mat/9906287; cond-mat/9911292.

\bibitem{21} R. J. Radwanski, {\it Physics of heavy-fermion phenomena} - Report CSSP-17/94
and CSSP-4/95 distributed in about 400 egzemplaires among leading
physicists at SCES-94 and SCES-95.

\bibitem{22} R. J. Radwanski and Z. Ropka, {\it Relativistic effects in the electronic structure for 3d
paramagnetic ions}, submitted in 1997 to Phys. Rev. Lett. LS
6925; available at ArXiv cond-mat/9907140).

\bibitem{23} R. J. Radwanski and Z. Ropka, {\it Strongly-Correlated Crystal Field
Approach to 3d oxides} (NovaScience USA) 2005; cond-mat/0404713.

\bibitem{24} Z. Ropka and R. J. Radwanski, Phys. Rev. B \textbf{67},
172401 (2001); R. J. Radwanski and Z. Ropka, cond-mat/0404713.

\bibitem{25} S. Noguchi, S. Kawamata, K. Okuda, H. Nojiri, and M.
Motokawa, Phys. Rev. B \textbf{66}, 094404 (2002).

\bibitem{26} M. A. Korotin, S. Yu. Ezhov, I. V. Solovyev, V. I.
Anisimov, D. I. Khomski, and G. A. Sawatzky, Phys. Rev. B
\textbf{54}, 5309 (1996).

\bibitem{27} R. J. Radwanski and Z. Ropka, Acta Phys. Pol. A
\textbf{97}, 963 (2000); cond-mat/0005471.

\bibitem{28}R. J. Radwanski and Z. Ropka, Physica B \textbf{345},
107 (2004); cond-mat/0306695.

\bibitem{29} R. J. Radwanski and Z. Ropka, J. Magn. Magn. Mater.
\textbf{272-276}, e259 (2004).

\bibitem{30} Z. Ropka, R. Michalski, and R. J. Radwanski, Phys. Rev. B \textbf{63},
172404 (2001).

\bibitem{31} R. J. Radwanski and Z. Ropka, {\it The atomic start description of NiO}, cond-mat/0503407.

\bibitem{32}A. Abragam and B. Bleaney, {\it Electron Paramagnetic
Resonance of Transition Ions} (Clarendon Press, Oxford) 1970, ch.
7.

\end{thebibliography}
\end{document}